\documentclass[12pt,a4]{article}
\usepackage{amsthm,amsmath,latexsym,multibox,amssymb,amsfonts}
\usepackage{graphicx,lscape,fancyhdr,array,stmaryrd}

\newcommand{\bep}{\begin{picture}}
\newcommand{\eep}{\end{picture}}

\newcommand{\FrameLikeAA}{{\unitlength=0.31mm\bep(10,120)(0,-20){%
\put(0,0){\Rect{1}{8}}\put(10,30){\Rect{1}{5}}%
\put(2,85){$p$}\put(12,85){$q$}%
}\eep}}

\newcommand{\FrameLikeA}{{\unitlength=0.31mm\bep(120,165)(0,-5){%
\put(0,0){\Rect{1}{14}}\put(10,30){\Rect{1}{11}}\put(20,80){\RectC{6}{3}{4}{2}{2}{1}}%
\put(2,145){$p$}\put(12,145){$q$}%
}\eep}}

\newcommand{\FrameLikeB}{{\unitlength=0.31mm\bep(120,165)(0,-5){%
\put(0,30){\Rect{1}{11}}\put(10,80){\RectC{6}{3}{4}{2}{2}{1}}%
\put(2,145){$q$}%
}\eep}}

\newcommand{\FrameLikeC}{{\unitlength=0.31mm\bep(120,165)(0,-5){%
\put(0,0){\Rect{1}{14}}\put(0,-10){\Rect{1}{1}}\put(10,80){\RectC{6}{3}{4}{2}{2}{1}}%
\put(2,145){$p$}%
}\eep}}

\newcounter{YoungHeight}\newcounter{YoungWidth}

\newcounter{Mul1}\newcounter{Mul2}\newcounter{Mul3}\newcounter{Mul4}
\newcounter{A0}\newcounter{A1}\newcounter{A2}\newcounter{A3}\newcounter{A4}\newcounter{A5}\newcounter{A6}
\newcounter{B0}\newcounter{B1}\newcounter{B2}\newcounter{B3}
\newcounter{C1}\newcounter{C2}\newcounter{C3}\newcounter{C4}\newcounter{C6}\newcounter{C7}
\newcounter{D1}\newcounter{D2}\newcounter{D3}\newcounter{D4}\newcounter{D5}
\newcounter{T0}\newcounter{T1}

\newcounter{TGR0}
\newcounter{R0}\newcounter{R1}\newcounter{R2}\newcounter{R3}
\newcounter{AR0}\newcounter{AR1}\newcounter{AR2}\newcounter{AR3}\newcounter{AR4}\newcounter{AR5}\newcounter{AR6}\newcounter{AR7}
\newcounter{Dotted0}\newcounter{Dotted1}\newcounter{Dotted2}\newcounter{Dotted3}

\newcounter{reptA}

\newlength{\txtHShift}

\newlength{\txtWidth}

\newcommand{\HalfLength}[2]{\setcounter{Mul1}{#1}\setcounter{Mul2}{#1}\addtocounter{Mul1}{\value{Mul2}}\addtocounter{Mul1}{\value{Mul2}}%
\addtocounter{Mul1}{\value{Mul2}}\addtocounter{Mul1}{\value{Mul2}}\setcounter{#2}{\value{Mul1}}}

\newcommand{\Add}[3]{\setcounter{#1}{#2}\addtocounter{#1}{#3}}

\newcommand{\Length}[1]{#10}

\newcommand{\shiftedText}[2]{{\hspace{#1}#2}}
\newcommand{\calcHShift}[1]{\settowidth{\txtWidth}{#1}\setlength{\txtHShift}{-0.5\txtWidth}}

\newcommand{\TextTop}[3]{{\calcHShift{#1}\HalfLength{#2}{T0}\Add{T1}{\Length{#3}}{-9}\put(\value{T0},\value{T1}){\shiftedText{\txtHShift}{#1}}}}

\newcommand{\Rect}[2]{\bep(\Length{#1},\Length{#2})\put(0,0){\line(1,0){\Length{#1}}}\put(0,0){\line(0,1){\Length{#2}}}%
\put(\Length{#1},\Length{#2}){\line(-1,0){\Length{#1}}}\put(\Length{#1},\Length{#2}){\line(0,-1){\Length{#2}}}\eep}

\newcommand{\RectB}[4]{{\Add{R0}{\Length{#2}}{\Length{#4}}\bep(\Length{#1},\value{R0})%
\put(0,0){\Rect{#3}{#4}}\put(0,\Length{#4}){\Rect{#1}{#2}}\eep}}

\newcommand{\RectC}[6]{{\Add{R0}{\Length{#2}}{\Length{#4}}\addtocounter{R0}{\Length{#6}}\bep(\Length{#1},\value{R0})%
\put(0,0){\Rect{#5}{#6}}\put(0,\Length{#6}){\RectB{#1}{#2}{#3}{#4}}\eep}}

\newcommand{\RectT}[3]{\bep(\Length{#1},\Length{#2})\put(0,0){\line(1,0){\Length{#1}}}\put(0,0){\line(0,1){\Length{#2}}}%
\put(\Length{#1},\Length{#2}){\line(-1,0){\Length{#1}}}\put(\Length{#1},\Length{#2}){\line(0,-1){\Length{#2}}}#3{#1}{#2}\eep}

\newcommand{\RectARow}[2]{{\bep(\Length{#1},10)\put(0,0){\RectT{#1}{1}{\TextTop{#2}}}\eep}}

\newcommand{\RectBRow}[4]{{\bep(\Length{#1},20)\put(0,0){\RectT{#2}{1}{\TextTop{#4}}}%
\put(0,10){\RectT{#1}{1}{\TextTop{#3}}}\eep}}

\usepackage{amsthm,amsmath,latexsym,multibox,amssymb,amsfonts,amsbsy}
\usepackage{graphicx,lscape,fancyhdr,array,stmaryrd,euscript,wrapfig}
\usepackage{cite}

\newcommand{\Sigm}{{\ensuremath{{\boldsymbol \sigma}_-}}}

\newcommand{\AlgebraFont}[1]{\mathfrak{#1}}

\newcommand{\lorentz}{{\ensuremath{\AlgebraFont{so}(d-1,1)}}}

\newcommand{\mls}{{\ensuremath{\AlgebraFont{so}(d-2)}}}

\newcommand{\pl}{\partial}

\newcommand{\fm}[1]{_{\mathbf{{#1}}}}

\newcommand{\be}{\begin{equation}}
\newcommand{\ee}{\end{equation}}
\newcommand{\bes}{\begin{split}}
\newcommand{\es}{\end{split}}
\newcommand{\bee}{\begin{eqnarray}}
\newcommand{\eee}{\end{eqnarray}}
\newcommand{\beee}{\begin{array}}
\newcommand{\bem}{\begin{multline}}
\newcommand{\eem}{\end{multline}}
\newcommand{\bec}{\begin{Comment}}
\newcommand{\ec}{\end{Comment}}

\newcommand{\Y}[1]{{\ensuremath{\mathbb{Y}(#1)}}}
\newcommand{\Ya}[1]{{\ensuremath{\mathbb{Y}[#1]}}}
\newcommand{\Ff}{\ensuremath{\mathbf{F}}}
\newcommand{\Yy}{\ensuremath{\mathbf{Y}}}

\newcommand{\Ss}{\ensuremath{\mathbf{S}}}

\newcommand{\fpl}{{\slash\!\!\!\partial}}

\newcommand{\DL}{{D}}

\newcommand{\BYoung}[4]{{{\smallpic{\RectBRow{#1}{#2}{${\scriptstyle #3}$}{${\scriptstyle #4}$}}}}}

\newcommand{\AYoung}[2]{{{\smallpic{\RectARow{#1}{${\scriptstyle #2}$}}}}}

\newcommand{\smallpic}[1]{{\unitlength=0.2mm#1}}

\newcommand{\scalp}[3]{{\left\langle\vphantom{#3}\vphantom{#2}{#1}\right|\left.#2\vphantom{#1}\vphantom{#3}\right|\left.\vphantom{#1}\vphantom{#2}{#3}\right\rangle}}

\begin{document}
\renewcommand{\thefootnote}{\fnsymbol{footnote}}
\begin{flushright}
\vspace{1mm}
\end{flushright}

\vspace{1cm}

\begin{center}
{\bf \Large  Frame-like Actions for Massless Mixed-Symmetry Fields in Minkowski space. Fermions } \vspace{1cm}

\textsc{E.D. Skvortsov\footnote{skvortsov@lpi.ru} and Yu.M.
Zinoviev\footnote{yurii.zinoviev@ihep.ru}}

\vspace{.7cm}

{\em${}^*$ P.N.Lebedev Physical Institute, Leninsky prospect 53, 119991, Moscow, Russia}

{\em${}^\dag$ Institute for High Energy Physics, Protvino, Moscow Region, 142280, Russia}

\end{center}

\vspace{0.5cm}
\begin{abstract}
A frame-like action for massless mixed-symmetry fermionic fields in Minkowski space is constructed.
The action is uniquely determined by gauge invariance.
\end{abstract}

\renewcommand{\thefootnote}{\arabic{footnote}}
\setcounter{footnote}{0}
\section{Introduction}\setcounter{equation}{0}
This paper fills a gap in the theory of mixed-symmetry gauge fields in Minkowski space. Namely, we extend the results of \cite{Skvortsov:2008sh, Zinoviev:2009vy} on frame-like Lagrangians to the case of massless fermionic fields of arbitrary symmetry type in Minkowski space of any dimension $d$. The term 'arbitrary symmetry' means that the field potential is a spin-tensor whose tensor indices could have the symmetry of any Young diagram, using 'mixed-symmetry' implies that the Young diagram contains more than just one row or one column.

In Minkowski space massless fields with the spin in arbitrary representation of the Wigner little algebra \mls{} were first considered by Labastida in the paper \cite{Labastida:1987kw}, where the equations of motion were suggested both for bosons and fermions. However, to find the action for mixed-symmetry fields turned out to be a complicated technical problem, solved in \cite{Labastida:1987kw} for bosons only. It was not until \cite{Campoleoni:2009gs} that the Lagrangian for arbitrary spin fermionic fields was constructed. The approach of Labastida is usually referred to as the metric-like approach, since the Labastida potentials, being the world tensors, are the analogs of the metric field $g_{\mu\nu}$.

Another approach to fields of any symmetry type dates back to \cite{Vasiliev:1980as, Lopatin:1987hz}, where the generalization of the vielbein/spin-connection variables of gravity was found for the massless spin-$s$ field, $s\geq2$. The approach that makes use of vielbein/spin-connection variables and their generalizations for fields of arbitrary symmetry type is referred to as frame-like one.

What is so inspiring about the frame-like approach is that it has a natural extension, the unfolded approach to field equations\cite{Vasiliev:1988xc, Vasiliev:1988sa}, where the generalized vielbein/spin-connection and certain their extensions appear automatically as the first fields in the unfolded system. The only example of an interacting field theory with fields of any tensor rank is the Vasiliev theory of totally symmetric gauge fields \cite{Vasiliev:1990en, Vasiliev:2003ev}. The point is that the Vasiliev theory is given in a form of nonlinear unfolded equations of motion. So we believe that investigating mixed-symmetry fields within the frame-like approach should give us some hints towards the interacting theory of mixed-symmetry fields.

The extension of the frame-like approach to fields of mixed-symmetry in Minkowski space was given first in \cite{Zinoviev:2003dd, Zinoviev:2003ix} for certain special cases. The general case of mixed-symmetry bosons and fermions was studied in \cite{Skvortsov:2008vs} at the level of equations of motion. The frame-like action for mixed-symmetry bosons was found in \cite{Skvortsov:2008sh}. Certain special cases in a more general setup of massless and massive fields in (anti)-de Sitter space were studied in \cite{Zinoviev:2008ve, Zinoviev:2009gh, Zinoviev:2009vy}.
Bosonic mixed-symmetry fields are more elaborated \cite{Labastida:1987kw, Metsaev:1993mj, Zinoviev:2002ye, Bekaert:2002dt, Zinoviev:2003dd,
Zinoviev:2003ix, Medeiros:2003dc, Alkalaev:2003hc, Alkalaev:2003qv, Metsaev:2005ar, Alkalaev:2006hq, Bekaert:2006ix, Buchbinder:2007ix, Skvortsov:2008sh,
Reshetnyak:2008gp, Skvortsov:2008vs, Campoleoni:2008jq, Zinoviev:2008ve,Boulanger:2008up, Boulanger:2008kw, Alkalaev:2008gi, Skvortsov:2009zu, Zinoviev:2009vy, Skvortsov:2009nv, Zinoviev:2009gh, Alkalaev:2009vm} as compared to mixed-symmetry fermions \cite{Labastida:1987kw, Moshin:2007jt, Skvortsov:2008sh, Zinoviev:2009vy, Campoleoni:2009gs}.

In section \ref{SSymmetric} we review the frame-like formulation for a
massless spin-$(s+\frac12)$ field. A short remark on totally antisymmetric fermionic fields is in section \ref{SOneColumn}. The general enough case of fermions whose tensor indices have the symmetry of a Young diagram with two columns is considered in section \ref{STwoColumn}. Fermions of any symmetry type are studied in section \ref{SGeneralCase}.

\section*{Notation}
Indices $\mu,\nu,...=0,...,(d-1)$ are the world indices of the base Minkowski space. $a, b,...=0,...,(d-1)$ are the indices of the fiber Lorentz algebra \lorentz, which are raised and lowered with constant metric $\eta_{ab}$. A differential form degree is indicated as a bold subscript,
\be\nonumber\omega\fm{q}\equiv\omega_{\mu_1...\mu_q}\,dx^{\mu_1}\wedge...\wedge dx^{\mu_q}\,.
\ee

A Young diagram is defined either by listing the lengths of its rows (row notation)
$\Y{s_1,s_2,...,s_n}$, $s_i$ being the length of the $i$-{th} row
($s_i\geq s_j$ for $i>j$), or by listing the heights of its columns [column notation] $\Ya{h_1,...,h_m}$, $h_j$
being the height of the $j$-th column ($h_i\geq h_j$ for $i>j$).

A group of symmetric indices is denoted by one letter, indicating the number of symmetric indices in round brackets, e.g. $a(k)\equiv a_1 a_2...a_k$. The operation of symmetrization is denoted by round brackets or by denoting the indices to be symmetrized by the same letter. The same index convention holds for antisymmetric indices with the replacement of round brackets by square brackets.
The operation of (anti)symmetrization is a sum over all {\it necessary}
permutations with the unit weight, e.g.,
\be\nonumber C^{ab}=C^{ba},\qquad\qquad
V^aC^{aa}\equiv V^{a_1}C^{a_2a_3}+V^{a_3}C^{a_1a_2}+V^{a_2}C^{a_3a_1}.
\ee That a tensor has  symmetry of some Young diagram $\Yy=\Y{s_1,s_2,...,s_n}$ refers only to
the permutation symmetry of its indices. An irreducible tensor of the
Lorentz algebra with symmetry $\Yy$ is totally
traceless in addition, i.e. contraction of any two its indices with $\eta_{ab}$ is zero. An irreducible spin-tensor with one spinor index is also defined by Young diagram that refers to the symmetry type of its tensor indices.

\section{Symmetric half-integer spin field}\label{SSymmetric}
A massless spin-$(s+\frac12)$ field can be described\cite{Fang:1978wz} by the Fang-Fronsdal potential $\phi_{(\mu_1...\mu_s):\alpha}$ that is totally symmetric in world indices $\mu_1...\mu_s$ and has a spinor index $\alpha$. The field equation, gauge transformations and algebraic constraints read as\footnote{The lagrangian equations of \cite{Fang:1978wz} have the form $G_{\mu_1...\mu_s}-\frac{s}2\Gamma_{(\mu_1}\Gamma^\nu G_{\nu\mu_2...\mu_s)}-\frac{s(s-1)}4\eta_{(\mu_1\mu_2}G^{\nu}_{\phantom{\nu}\nu\mu_3...\mu_s)}=0$, $G_{\mu_1...\mu_s}=\fpl\phi_{\mu_1...\mu_s}-s\pl_{(\mu_1}\Gamma^{\nu}\phi_{\nu\mu_2...\mu_s)}=0$, which is equivalent to (\ref{FlatMSFronsdalFermionic}).}
\be\label{FlatMSFronsdalFermionic}\begin{split}
    &\fpl\phi_{\mu_1...\mu_s}-s\pl_{(\mu_1}\Gamma^{\nu}\phi_{\nu\mu_2...\mu_s)}=0,\\
    &\delta \phi_{\mu_1...\mu_s}=\pl_{(\mu_1}\xi_{\mu_2...\mu_s)},\\
    &\Gamma^\nu\Gamma^\rho\Gamma^\lambda\phi_{\nu\rho\lambda\mu_4...\mu_s}=0,\qquad \Gamma^\nu\xi_{\nu\mu_2...\mu_{s-1}}=0,
\end{split}\ee
where ${\Gamma_m}^{\alpha}_{\phantom{\alpha}\beta}$ are the gamma-matrices of \lorentz{} and we omit spinor indices in most cases.
The triple $\Gamma$-trace constraints, which are weaker than the irreducibility of $\phi_{\mu_1...\mu_s:\alpha}$, are needed in order to have a Lagrangian.

Within the frame-like approach the analogue of the vielbein (frame) for a massless spin-$(s+\frac12)$ field is given\footnote{In four dimension this construction was introduced in \cite{Vasiliev:1980as} and in arbitrary dimension in \cite{Vasiliev:1987tk}, the lagrangian that is considered in this paper is from \cite{Zinoviev:2008ze}.} by a degree-one differential form $e\fm{1}^{a(s-1):\alpha}$ that is a rank-$(s-1)$ irreducible symmetric spin-tensor of the Lorentz algebra that acts on fiber indices of the vielbein, i.e.
\begin{align} &e\fm{1}^{a(s-1)}\equiv e_\mu^{a_1...a_{s-1};\alpha}\, dx^\mu, && {\Gamma_m}^{\alpha}_{\phantom{\alpha}\beta}\, e_\mu^{m a_2...a_{s-1};\beta}\equiv0.\label{SpinSFrameA}\end{align}  As a consequence of (\ref{SpinSFrameA}) the vielbein is traceless in tensor indices $a_1...a_{s-1}$.

Within the frame-like approach the background Minkowski space is described by background vielbein $h^a_\mu dx^\mu$ and spin-connection $\varpi^{a,b}_\mu dx^\mu=-\varpi^{b,a}_\mu dx^\mu$, which obey
\begin{align}d h^a+\varpi^{a,}_{\phantom{a,}c}\wedge h^c=0,\label{FlatnessA}\\ d\varpi^{a,b}+\varpi^{a,}_{\phantom{a,}c}\wedge\varpi^{c,b}=0.\label{FlatnessB}\end{align}
A particular solution is given by Cartesian coordinates $h^a_\mu=\delta^a_\mu$, $\varpi^{a,b}_\mu=0$. In what follows we do not need any explicit solution of (\ref{FlatnessA})-(\ref{FlatnessB}), making the Lagrangians valid in any coordinates. The Lorentz covariant derivative $\DL$ is defined as $\DL=d+\varpi$, where $d$ is the exterior derivative $d\equiv dx^\mu\, \pl_\mu$.

The anzats for the action consists of two terms, if only exterior derivative $d$ and exterior product $\wedge$ are allowed to be used\footnote{We omit everywhere the wedge symbol $\wedge$ henceforth.},
\begin{align}S=\int \left[{{{\bar{e}}\fm{1}}}{}^{ua(s-2)}\Gamma^v \DL {e\fm{1}}^{w}_{\phantom{w}a(s-2)}-\frac16{\bar{e}\fm{1}}{}^{a(s-1)}\Gamma^{uvw} \DL {e\fm{1}}{}_{a(s-1)}\right]E_{uvw},\end{align}
where $E_{u[k]}$ and $\Gamma^{b[N]}$ are defined as
\begin{align}E_{u[k]}&\equiv\epsilon_{u[k]b_1...b_{d-k}}h^{b_1}...h^{b_{d-k}},\\
\Gamma^{b[N]}&\equiv \frac1{N!}\Gamma^{[b_1}...\Gamma^{b_N]},\end{align}  where $\epsilon_{u_1...u_d}$ is the totally antisymmetric invariant tensor of the Lorentz algebra.

The relative coefficient in the Lagrangian is determined by requiring the action be gauge invariant under
\be\delta e\fm{1}^{a(s-1)}=\DL \xi\fm{0}^{a(s-1)} + h_m \xi\fm{0}^{a(s-1),m},\ee
where $\xi\fm{0}^{a(s-1):\alpha}$ is a gauge parameter for the frame $e\fm{1}^{a(s-1):\alpha}$ and it is just the fiber version of the Fang-Fronsdal parameter
$\xi_{\mu_1...\mu_{s-1}:\alpha}$. There is also a shift (Stueckelberg) gauge symmetry with $\xi\fm{0}^{a(s-1),m}$ that as a fiber tensor has the symmetry of \smallpic{\RectBRow{5}{1}{$\scriptstyle s-1$}{}} and is irreducible as a spin-tensor.

The variation with respect to $\xi\fm{0}^{a(s-1)}$ is a total derivative. To check that the $\xi\fm{0}^{a(s-1),m}$-variation also cancels one has to use the following useful identities\footnote{The hatted index is to be omitted.}
\begin{align}\label{MSIdentityVielbeins}h^c E_{u_1...u_k}=\frac1{d-k+1}\sum_{i=1}^{i=k}(-)^{i+k}\delta^c_{u_i}E_{u_1...\widehat{u}_i...u_k},\\
\label{MSIdentityGammaA}\Gamma^a\Gamma^{b_1...b_N}=
\sum_{i=1}^{i=N}(-)^{i+1}\eta^{ab_i}\Gamma^{b_1...\widehat{b}_i...b_N}+\Gamma^{ab_1...b_N},\end{align}
and hence for an irreducible spin-tensor $\xi^{a(s-1),m}$ one has
\be\label{MSIdentityGammaB}{\xi}^{a(s-1),}_{\phantom{a(s-1),}m} \Gamma^{mb_1...b_{N-1}}=\sum^{i=N-1}_{i=1}(-)^i{\xi}^{a(s-1),{b_i}}\Gamma^{b_1...\widehat{b}_i...b_{N-1}}.\ee

The meaning of the shift symmetry is two-fold. Firstly, it compensates for a larger number of components in the vielbein. Indeed, with the help of the background frame field we can transform any differential form with fiber spin-tensor indices into fully fiber spin-tensor and then decompose it into irreducible spin-tensors. For $e\fm{1}^{a(s-1)}$ it results in \be e_\mu^{a(s-1)}h^{b\mu}\sim\BYoung{5}{1}{ s-1}{}\oplus\AYoung{6}{s}\oplus\AYoung{5}{s-1}\oplus\AYoung{4}{s-2},\ee where the first component can be gauged away by the $\xi\fm{0}^{a(s-1),m}$ transformations, the rest of components matching the content of the Fang-Fronsdal field $\phi_{\mu_1...\mu_s:\alpha}$, whose first and second $\Gamma$-traces do not vanish.

Secondly, it suggests that there is a field $\omega\fm{1}^{a(s-1),m}$, which is a gauge field for the $\xi\fm{0}^{a(s-1),m}$-symmetry. For the spin-two case, $\omega\fm{1}^{a,b}$ is just a spin-connection and has a plain geometric meaning. If a field potential has the symmetry of a Young diagram with more than two columns then there are fields that follow generalized spin-connection. These fields are called {\it extra fields} and they are not needed at the free level.

\section{One-column field}\label{SOneColumn}
Within the metric-like approach totally antisymmetric fermionic fields were studied in \cite{Buchbinder:2009pa} and \cite{Zinoviev:2009wh}. The field potential is an antisymmetric spin-tensor $\phi_{\mu_1...\mu_p:\alpha}$ with the gauge transformations of the form
\be\delta\phi_{\mu_1...\mu_p}=\pl_{[\mu_1}\xi_{\mu_2...\mu_p]}.\ee
Both the field potential and gauge parameter are not subjected to any $\Gamma$-trace conditions.

According to \cite{Skvortsov:2008vs}, the frame-like extension for a spin $\Ss=\Ya{p}$ field is given by a vielbein $e\fm{p}^{\alpha}$, which is a $p$-form with one fiber spinor index. The gauge transformation law is just
\be \label{TwoColumnGaugeA}\delta e\fm{p}^{\alpha}=\DL \xi\fm{p-1}^{\alpha}.\ee
There is no shift symmetry here because in this degenerate case the vielbein has the same number of components as the field potential has.

The anzats for the action consists of one term only \be\label{TwoColumnAction}
S=\int\scalp{{\bar{e}\fm{p}}}{\Gamma^{u[2M-1]}}{\DL{e\fm{p}}}
E_{u[2p+1]},\ee where $M=p-q+1$ and the brackets $\langle \bar{e}|\Gamma|\DL e\rangle$ just stress that all spinor indices are contracted. Evidently, the action is gauge invariant.

We see that it is not possible to write down the action if $2p+1>d$, which is exactly the bound found recently in \cite{Buchbinder:2009pa} and \cite{Zinoviev:2009wh}. If $2p+1>d$ it is still possible to write a lot of terms for the anzats with metric-like field $\phi_{\mu[p]:\alpha}$, which results in a zero Lagrangian after requiring its gauge variation to vanish. We see that certain features, which require special consideration within the metric-like approach, are automatically included in the frame-like one.

\section{Two-column field}\label{STwoColumn}
\begin{wrapfigure}{r}{2cm} \Ss=\parbox{1cm}{\FrameLikeAA}
\end{wrapfigure}Let us now turn to a more complicated case of fields whose spin $\Ss$ is defined by a Young diagram with two columns $\Ss=\Ya{p,q}$. We will not present the results for the metric-like approach\cite{Campoleoni:2009gs}, restricting ourselves to the frame-like one. In fact, within the frame-like approach general mixed-symmetry fields turn out to be as complicated as two-column fields. This is because the rest of indices corresponding to the third and other columns are mutually contracted in the action. Hence the presence of indices beyond the second column does not affect any coefficients coming from the variation of the action.

According to \cite{Skvortsov:2008vs}, the frame-like extension for a spin $\Ss=\Ya{p,q}$ field is given by a vielbein $e\fm{p}^{u[q]}$, which is a $p$-form antisymmetric in $q$ fiber indices, and spin-connection $\omega\fm{q}^{u[p+1]}$, which is a $q$-form antisymmetric in $(p+1)$ fiber indices\footnote{Let us note that the fiber spinor index is implicit.}. Despite the fact that $\omega\fm{q}^{u[p+1]}$ drops out of the action, its gauge parameter $\xi\fm{q-1}^{u[p+1]}$ is needed to have the correct number of propagating degrees of freedom.
The gauge transformations
\be \label{TwoColumnGaugeA}\delta e\fm{p}^{u[q]}=\DL \xi\fm{p-1}^{u[q]}+ \overbrace{h_m...h_m}^M \xi\fm{q-1}^{u[q]m[M]},\ee
guarantee\cite{Skvortsov:2008vs} that the vielbein $e\fm{p}^{u[q]:\alpha}$ effectively reduces to the potential $\phi_{\mu[p],\nu[q]:\alpha}$, the rest of components being pure gauge because of the $\xi\fm{q-1}^{u[p+1]:\alpha}$ symmetry.

The most general anzats for the action has the form \be\label{TwoColumnAction}
S=\sum_{k=0}^{k=q}\alpha_k\int
\scalp{{\bar{e}\fm{p}}{}^{u[q-k]b[k]}}{\Gamma^{u[2M+2k-1]}}{\DL{e\fm{p}}^{u[q-k]}_{\phantom{u[q-k]}b[k]}}
E_{u[2p+1]},\ee
where the brackets $\langle \bar{e}|\Gamma|\omega\rangle$ stress that all spinor and tensor indices of $e$, $\Gamma$ and $\omega$ are contracted in an appropriate way, with only $(2p+1)$ antisymmetric indices left on the outside to be contracted with $E_{u[2p+1]}$.

The action is invariant up to surface terms under the differential part of the gauge transformations. To take the variation with respect to the shift gauge symmetry one needs
an extension of the identities (\ref{MSIdentityVielbeins}) and (\ref{MSIdentityGammaB})
\begin{align}
&h^{c_1}...h^{c_m}E_{u[N]}=m!(-)^{m(N+1)}\frac{(d-N)!}{(d-N+m)!}\,\delta^{c[m]}_{u[m]}E^{\phantom{c[m]}}_{u[N-m]},&&\label{MSIdentityVielbeinsX}\\
&{\xi}_{c_1...c_m}\Gamma^{c[m]b[N]}=m!(-)^{\frac{m(m+1)}2}{\xi}^{b[m]}\Gamma^{b[N-m]},\label{MSIdentityGammaX}
\end{align}
where the latter identity holds true provided that the spin-tensor is $\Gamma$-traceless, ${\xi}^{c[m-1]b}\Gamma_b\equiv0$. It is also important that the antisymmetrization includes all necessary permutations only. After some simple algebra one obtains
\begin{align*}&\delta S\sim\sum_{k=0}^{k=q}\alpha_k\sum_{n=0}^{n=\min(M,q-k)}F_{k,n}(-)^{\sigma_{k,n}}\times\\&
\int\scalp{{\bar{\xi}\fm{q-1}}{}^{u[q-k-n]b[k+n]}}{\Gamma^{v[2k+2n-1]}}
{\DL{e\fm{p}}^{w[M+q-k-n]\phantom{b[k+n]}}_{\phantom{w[M+q-k-n]}b[k+n]}}E_{u[q-k-n]v[2k+2n-1]w[M+q-k-n]},
\end{align*}
where
\begin{align*}F_{k,n}&=\frac{(q-k)!(2M+2k-1)!}{(q-k-n)!n!(M-n)!(2k+2n-1)!},\\
(-)^{\sigma_{k,n}}&=(-)^{\frac12(M-n)(M+2k-n+1)+q(q-n)}.\end{align*}
The condition for the variation to vanish is thus
\be\sum_{k=0,\,k+n=m}^{k=\min(M,q-k)}\alpha_k F_{k,n}(-)^{\sigma_{k,n}}=0\qquad \mbox{for any} \quad m\in[1,q].\ee
Note that for $k=n=0$ the variation is zero identically. The general solutions is \be\label{TwoColumnCoef}\alpha_k=\alpha(-)^{\frac12k(k+2q+1)}\frac{(M+k-1)!}{(q-k)!(2M+2k-1)!k!}\ee
with $\alpha$ being an arbitrary constant depending only on $p,q$, i.e. on the spin.

If we define the field strength
\be R\fm{p+1}^{u[q]}=\DL e\fm{p}^{u[q]}+ (-)^{p-q}h_m...h_m\, \omega\fm{q}^{u[q]m[M]},\ee
which is gauge invariant under (\ref{TwoColumnGaugeA}) and
\be \label{TwoColumnGaugeA}\delta \omega\fm{q}^{u[p+1]}=\DL \xi\fm{q-1}^{u[p+1]}+...,\ee
where $...$ stands for the shift symmetry associated to extra fields, which are beyond $\omega\fm{q}^{u[p+1]}$ and are not needed for the purpose of constructing a free Lagrangian. Then there is an equivalent form for the action
\be
S=\sum_{k=0}^{k=q}\alpha_k\int
\scalp{{\bar{e}\fm{p}}{}^{u[q-k]b[k]}}{\Gamma^{u[2M+2k-1]}}{{R\fm{p+1}}^{u[q-k]}_{\phantom{u[q-k]}b[k]}}
E_{u[2p+1]},\ee where $\alpha_k$ are determined by requiring the action not to contain $\omega\fm{q}^{u[p+1]}$. Note that for bosons the extra fields decouple out of the action automatically. By contrast, for fermions one has to adjust certain coefficients for the extra field decoupling condition to hold true.

\section{Arbitrary-spin massless fermionic fields}\label{SGeneralCase}
The result of \cite{Skvortsov:2008vs, Skvortsov:2008sh} was that the frame-like approach can be generalized to massless fields having arbitrary symmetry type. Some special cases at the level of Lagrangian were considered in \cite{Zinoviev:2003dd, Zinoviev:2003ix, Zinoviev:2008ve, Zinoviev:2009gh, Zinoviev:2009vy}.

In the general case one has to deal with differential forms of any degree that carry fiber tensor/spinor indices of the Lorentz algebra and are irreducible as fiber spin-tensors. Let us denote the symmetry type of the fiber tensor indices by superscript Young diagram, e.g. a $q$-from $\omega$ that is an irreducible fiber spin-tensor whose tensor indices have the symmetry of Young diagram $\Yy$ is denoted as $\omega\fm{q}^\Yy$.

Let $\Ss=\Ya{h_1,h_2,...,h_n}\equiv\Y{s_1,...,s_p}$ be the spin Young diagram, where depending on the problem it is convenient to enumerate either columns $h_i$ or the rows $s_i$. It is useful to single out the first two columns of $\Ss$,
denoting $p=h_1$, $q=h_2$, so that $\Ss=\Ya{p,q,h_3,...,h_n}$. According to \cite{Skvortsov:2008vs}, the vielbein and spin-connection for a spin-$\Ss$ field are $e^{\Ff_0}\fm{p}$ and $\omega^{\Ff_1}\fm{q}$, where $\Ff_0$ and $\Ff_1$ are given by

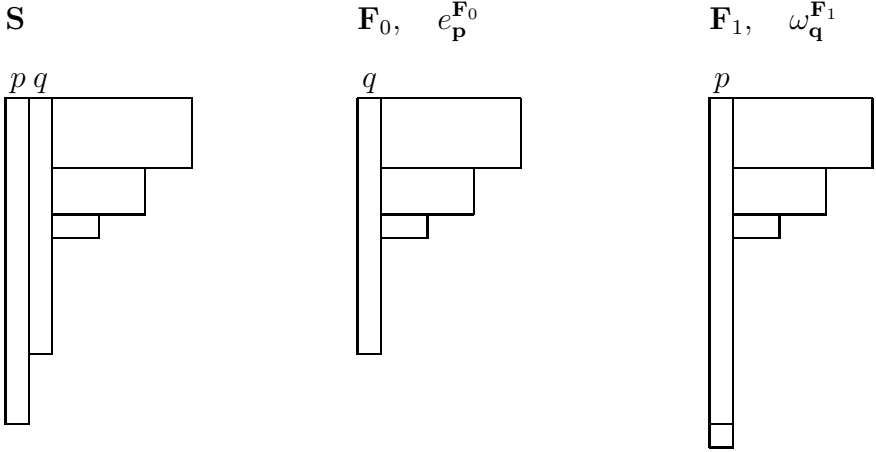
\begin{align*}& \Ss && \Ff_0, \quad e^{\Ff_0}\fm{p} && \Ff_1, \quad \omega^{\Ff_1}\fm{q}\\
&\FrameLikeA && \FrameLikeB&&\FrameLikeC\end{align*}
\begin{align*}
\Ff_0&=\Ya{q,h_3,...,h_{s_1}}, && \Ff_1=\Ya{p+1,h_3,...,h_{s_1}}.
\end{align*}
With all indices written explicitly the vielbein has the form
\be e^{\Ff_0}\fm{p}\equiv e^{a(s_1-1),b(s_2-1),...,c(s_q-1):\alpha}_{\mu_1...\mu_p}\,dx^{\mu_1}...dx^{\mu_p},
\ee where the symmetric basis for mixed-symmetry tensors was used, i.e. the indices are separated into the groups of symmetric ones. The Labastida potential $\phi^{a_1(s_1),...,a_p(s_p):\alpha}$ is the maximally symmetric component of the frame field
\be\phi^{a(s_1),b(s_2),...,u(s_p):\alpha}=e^{a(s_1-1),b(s_2-1),...,u(s_p-1):\alpha|ab...u}, \ee
where the form indices were converted to the fiber
ones according to \be e^{a(s_1-1),...,c(s_p-1):\alpha|u_1...u_p}=e^{a(s_1-1),...,c(s_p-1):\alpha}_{\mu_1...\mu_p}h^{\mu_1 u_1}...h^{\mu_p u_p}.\ee

\noindent The gauge transformations for $e^{\Ff_0}\fm{p}$ have the form
\begin{align}
\delta e^{a(s_1-1),...,u(s_q-1)}\fm{p}&=\DL \xi^{a(s_1-1),...,u(s_q-1)}\fm{p-1}+\overbrace{h_m...h_m}^{M}\xi^{a(s_1-1),...,u(s_q-1),\overbrace{\scriptstyle m,...,m}^{M}}\fm{q-1}.\label{FieldStrengthA}
\end{align}
It is easy to see that the shift symmetry does not affect the Labastida potential hidden inside the vielbein.

Let the operator contracting the indices in (\ref{FieldStrengthA}) with the background vielbein $h^a$ be denoted as $\Sigm$, then the field strength and gauge transformations read as
\begin{align}\label{ActionFieldStrengthA}
R^{\Ff_0}\fm{p+1}&=\DL e^{\Ff_0}\fm{p\phantom{-1}}+\Sigm(\omega^{\Ff_1}\fm{q})(-)^{p-q}, \\
\delta e^{\Ff_0}\fm{p}&=\DL \xi^{\Ff_0}\fm{p-1}+\Sigm(\xi^{\Ff_1}\fm{q-1}),\label{MSActionGaugeA}\\
\delta \omega^{\Ff_1}\fm{q}&=\DL \xi^{\Ff_1}\fm{q-1}+...,\label{MSActionGaugeB}
\end{align}
where $...$ denotes the contribution of gauge parameters of the extra connections, which can be
introduced but do not contribute to the free action.

\subsection{Action}
By looking at the action for mixed-symmetry bosons \cite{Skvortsov:2008sh, Zinoviev:2009gh}, to the action for fermions \cite{Zinoviev:2009vy} whose spin is defined by a Young diagram with two rows and to (\ref{TwoColumnAction}) we see that it is more natural to have explicit antisymmetrization over certain fiber indices, which seem to play the role of fiber differential forms.

To have explicit antisymmetry in certain fiber indices the mixed basis for mixed-symmetry fields is defined below.
Namely, taking some mixed-symmetry tensor, say $C^{\Yy}$ with $\Yy=\Y{k_1,...,k_n}$, in the symmetric basis
\be \label{MixedBasisA} C^{a(k_1),b(k_2),...,c(k_n)}.\ee
We then take one index from each of the group of symmetric indices \be \nonumber C^{ua(k_1-1),ub(k_2-1),...,uc(k_n-1)}\ee and antisymmetrize them
\be \label{MixedBasisC} C^{u[n];a(k_1-1),b(k_2-1),...,c(k_n-1)}\equiv C^{[u|a(k_1-1),|u|b(k_2-1),...,|u]c(k_n-1)}.\ee
As it can be easily verified with the help of Young symmetry properties, the resulting tensor has the symmetry of $\Y{k_1-1,...,k_n-1}$ in indices $a(k_1-1),...,c(k_n-1)$, it is by construction antisymmetric in $u_1...u_n$ and the antisymmetrization of any $n+1$ indices vanishes identically. Note that if some of $k_i$'s are equal to $1$, the corresponding groups of indices in (\ref{MixedBasisC}) will be empty. It is also straightforward to check that it is indeed a basis and hence there is an inverse map from (\ref{MixedBasisC}) to (\ref{MixedBasisA}), which up to some factor is given by the symmetrization of each of $u_1...u_n$ with one of the group of symmetric indices.

Let the spin of a field be given by some Young diagram $\Ss=\Ya{p,q,h_3,...,h_n}$. The indices that do not 'belong' to the first column of the vielbein $e^{\Ff_0}\fm{p}$ and spin-connection $\omega^{\Ff_1}\fm{q}$ are just mutually contracted in the action. Moreover, in any computations, for example when taking the variation of the action, the indices that are beyond the first column remain blind. There are two antisymmetric objects to be contracted with the fiber indices, these are $\epsilon_{u_1...u_d}$ and $\Gamma^{u[k]}$. Any contractions with these two objects that involve indices beyond the first column can be expressed as certain permutations of indices applied to the terms in which all the indices contracted with $\epsilon_{...}$ and $\Gamma^{...}$ do correspond to the first column. Actually, as shown in \cite{Skvortsov:2010at} the Young and trace conditions for the indices beyond the first column are not so important and can be relaxed, so that the action will describe a reducible set of fields, which generalizes the results of \cite{Sorokin:2008tf} for spin-$s$ fields.

The frame field $e^{\Ff_0}\fm{p}$ taken in the mixed basis reads as
\be e^{\Ff_0}\fm{p}\equiv e^{u[q];a(s_1-2),b(s_2-2),...,u(s_n-2):\alpha}\fm{p}.\ee
In the action the fiber indices $a(s_1-2),...$ are just mutually contracted
\be\label{Contraction}\scalp{{\bar{e}\fm{p}}{}^{u[q-k]b[k];a(s_1-2),...}}{\Gamma^{v[2M+2k-1]}}
{{\DL e\fm{p}}{}^{w[q-k]\phantom{b[k]};}_{\phantom{w[q-k]}b[k]\phantom{;}a(s_1-2),...}}
\ee this will remain the case when
taking any variation of the action. With the convention that we will not write the indices $a(s_1-2),...$ explicitly, the action has the same form (\ref{TwoColumnAction}) as for two-column fields. The coefficients $\alpha_k$ are of course also the same, given by (\ref{TwoColumnCoef}).

The subtle point for mixed basis could be that one should substitute the gauge transformations of the form \be\delta
{e\fm{p}}^{u[q];a(s_1-2),...}=h_m...h_m{\xi\fm{q-1}}^{u[q]m[M];a(s_1-2),...}\,.\label{GaugeTransformationAlgebraic}\ee As it is written the {\it r.h.s} in the latter expression does not have definite Young symmetry properties. Despite this fact one is safe to plug (\ref{GaugeTransformationAlgebraic}) into the action because the contraction of it with a tensor having definite Young symmetry of $\Ff_0$ projects out the components with symmetry other than $\Ff_0$.

We have constructed the frame-like actions for mixed-symmetry fermionic fields, whose tensor part is defined by arbitrary Young diagram.
As it is always the case the actions for fermionic fields are more complicated than for bosons, still treatable though.

\section*{Acknowledgements}
The work of E.D.S was supported in part by grant RFBR No. 08-02-00963 and grant UNK-FIAN. E.D.S. is grateful to M.A.Vasiliev for useful remarks on an early version of this paper.

\providecommand{\href}[2]{#2}\begingroup\raggedright\endgroup
\end{document}